\documentclass[twocolumn]{aastex631}
\usepackage{array}
\usepackage{multirow}
\newcolumntype{P}[1]{>{\centering\arraybackslash}p{#1}}
\usepackage{graphicx}	
\usepackage{amsmath}	
\usepackage{amssymb}	
\usepackage{comment}
\usepackage{enumitem}
\usepackage{newtxtext,newtxmath}
\usepackage{nicefrac}
\usepackage{natbib}

\newcommand{\angstrom}{\mbox{\normalfont\AA}}

\usepackage{amsmath} 

\graphicspath{{./}{figures/}}

\accepted{June 29, 2023 to ApJ}

\shorttitle{Continuum Reverberation Mapping of Mrk~876}
\shortauthors{Miller et al.}

\begin{document}

\title{Continuum Reverberation Mapping of Mrk~876 Over Three Years With Remote Robotic Observatories}
\author[0000-0001-8475-8027]{Jake A. Miller}
\affil{Wayne State University, Department of Physics \& Astronomy, 666 W Hancock St, Detroit, MI 48201, USA}

\author[0000-0002-8294-9281]{Edward M. Cackett}
\affiliation{Wayne State University, Department of Physics \& Astronomy, 666 W Hancock St, Detroit, MI 48201, USA}

\author[0000-0002-2908-7360]{Michael R. Goad}
\affiliation{School of Physics and Astronomy, University of Leicester, University Road, Leicester, LE1 7RH, UK}

\author[0000-0003-1728-0304]{Keith Horne}
\affiliation{SUPA Physics and Astronomy, University of St. Andrews, Fife, KY16 9SS Scotland, UK}

\author[0000-0002-3026-0562]{Aaron J. Barth}
\affiliation{Department of Physics and Astronomy, 4129 Frederick Reines Hall, University of California, Irvine, CA, 92697-4575, USA}

\author[0000-0003-0607-1136]{Encarni Romero-Colmenero}
\affiliation{South African Astronomical Observatory, P.O. Box 9, Observatory 7935, Cape Town, South Africa} 
\affiliation{Southern African Large Telescope Foundation, P.O. Box 9, Observatory 7935, Cape Town, South Africa}

\author[0000-0002-9113-7162]{Michael Fausnaugh}
\affiliation{Department of Astronomy, The Ohio State University, 140 W 18th Ave,
Columbus, OH 43210, USA}

\author[0000-0001-9092-8619]{Jonathan Gelbord}
\affiliation{Spectral Sciences Inc., 4 Fourth Avenue, Burlington, MA 01803, USA}

\author[0000-0003-0944-1008]{Kirk T. Korista}
\affiliation{Department of Physics, Western Michigan University, 1120 Everett Tower, Kalamazoo, MI 49008-5252, USA}

\author[0000-0001-8391-6900]{Hermine Landt}
\affiliation{Centre for Extragalactic Astronomy, Department of Physics, Durham University, South Road, Durham DH1 3LE, UK}

\author[0000-0002-8460-0390]{Tommaso Treu}
\affiliation{Department of Physics and Astronomy, University of California, Los Angeles, CA 90095, USA}

\author[0000-0003-2662-0526]{Hartmut Winkler}
\affiliation{Department of Physics, University of Johannesburg, P.O. Box 524, 2006 Auckland Park, South Africa}


\begin{abstract}
Continuum reverberation mapping probes the sizescale of the optical continuum-emitting region in active galactic nuclei (AGN). Through 3 years of multiwavelength photometric monitoring in the optical with robotic observatories, we perform continuum reverberation mapping on Mrk~876. All wavebands show large amplitude variability and are well correlated. Slow variations in the light curves broaden the cross-correlation function (CCF) significantly, requiring detrending in order to robustly recover interband lags. We measure consistent interband lags using three techniques (CCF, JAVELIN, PyROA), with a lag of around 13~days from $u$ to $z$. These lags are longer than the expected radius of 12~days for the self-gravitating radius of the disk. The lags increase with wavelength roughly following $\lambda^{4/3}$, as would be expected from thin disk theory, but the lag normalization is approximately a factor of 3 longer than expected, as has also been observed in other AGN. The lag in the $i$ band shows an excess which we attribute to variable H$\alpha$ broad-line emission. A flux-flux analysis shows a variable spectrum that follows $f_\nu \propto \lambda^{-1/3}$ as expected for a disk, and an excess in the $i$ band that also points to strong variable H$\alpha$ emission in that band. 
\end{abstract}


\section{Introduction} \label{sec:intro}
It is now known that most, if not all, galaxies host a supermassive black hole (SMBH) in their nuclei \citep{1998Natur.395A..14R}. Active galactic nuclei (AGN) occur when material falls onto this SMBH. The conservation of angular momentum is expected to lead to the formation of an accretion disk. Most of these disks are too far away to be spatially resolved directly with current instruments, so indirect techniques are needed to observe and understand the AGN system. By measuring how the irradiated areas surrounding the SMBH respond to changes from the ionizing source, one can use time lags between different wavelength bands to measure the size scale. This technique is called reverberation mapping \citep{1982ApJ...255..419B, 1993PASP..105..247P, 2014SSRv..183..253P} and has been successfully carried out on a number of different AGN. Many regions of the AGN can be probed in this way, from the innermost areas of the accretion disk out to the dusty torus. For more information on reverberation mapping, a review can be found in \cite{2021iSci...24j2557C}. 

Time lags between the continuum at different wavelengths are expected for reverberation of the accretion disk. In the lamppost model \citep{2002ApJ...567...85N, 2003A&A...400..437C, 2007MNRAS.380..669C}, X-rays from an ionizing source located above the SMBH irradiate the accretion disk, where they are reprocessed and re-emitted at longer wavelengths depending on the temperature of the disk where they land. The disk is hotter closer to the SMBH, producing strong ultraviolet and continuum emission, while further and cooler regions of the disk predominantly produce the optical continuum. This model predicts that the X-rays should drive and lead the variability seen at longer wavelengths. Inner regions of the accretion disk should then see variations first, followed by regions further away from the SMBH. The measured time lag is then assumed to be dominated by the light travel time between different regions of the disk. 
A geometrically thin, optically thick accretion disk \citep{1973A&A....24..337S} is expected to have a disk radial temperature profile of $T(R) \propto R^{-3/4}$. Since wavelength maps to temperature via the Wien displacement law and time lag to radius this gives wavelength dependent lags, $\tau(\lambda)$, which should scale with wavelength as $\tau(\lambda)\propto\lambda^{4/3}$ \citep{1999MNRAS.302L..24C}. This relationship has been observed in previous continuum reverberation mapping studies, both in focused multi-instrument campaigns \citep[e.g.,][]{2015ApJ...806..129E, 2016ApJ...821...56F, 2020MNRAS.498.5399H, 2021ApJ...922..151K} and from AGN survey studies \citep{2017ApJ...836..186J, 2018ApJ...862..123M, 2022ApJ...929...19G}. However, using a sample of over 9000 quasars from the SDSS Southern Survey \cite{2022MNRAS.512..899W} find a bluer spectrum, $\tau(\lambda)\propto\lambda^{5/7}$, when possible Small Magellanic Cloud-like dust is considered. 

Recent studies have shown some predictions of the lamppost model are not always valid, at least not for all AGN. Many AGN show a weaker correlation, or even a de-coupling, between the X-ray and the ultraviolet/optical light curves \citep[e.g.,][]{2017ApJ...840...41E, 2019ApJ...870..123E, 2019ApJ...870...54M, 2021ApJ...922..151K}. Some AGN have trends indicative of systems more complicated than the lamppost model. Several studies have shown that there appears to be an incoming slow-moving lag, where the optical leads the ultraviolet and X-rays over the span of 100s to 1000s of~days \citep[e.g.,][]{2009MNRAS.394..427B,2020MNRAS.498.5399H,2022MNRAS.tmp..982N}. This implies that on longer timescales it is changes in the accretion disk's accretion flow that drive the variability, not an ionizing source. There is also evidence for obscuring elements, such as disk winds, that could exist between the ionizing source and the accretion disk, severing their direct connection \citep[e.g.,][]{2021ApJ...922..151K}. Previous reverberation mapping campaigns have found lags that indicate accretion disks $\sim$3 times larger than expected \citep[e.g.,][]{2014ApJ...788...48S, 2015ApJ...806..129E, 2016ApJ...821...56F, 2017ApJ...836..186J}. However, certain physical models have been created \citep{2019ApJ...879L..24K,kammoun21a,kammoun21b} that give lags consistent with what is observed. More studies involving long-term multi-wavelength campaigns spanning several years are needed to understand variability on different timescales in AGN. 


 Mrk~876 is a prime target for such a study. Its location in the sky allows it to be observed by northern ground-based robotic observatories nearly year-round. It has a history of reliable variability, and has been studied in several other independent campaigns \citep{2000ApJ...533..631K, 2013ApJ...767..149B, 2022MNRAS.511.3005J, Landt2022}. From 2016 to 2019 it was observed by the Las Cumbres Observatory, the Liverpool Telescope, and the Dan Zowada Memorial Observatory. By combining these campaigns, a long-term look into the variability patterns and reverberation lags in Mrk~876 can be performed. In Section~\ref{sec:methods} we describe the data reduction and analysis, in Section~\ref{sec:results} we present the results from time lag and spectral variability analysis, and in Section~\ref{sec:discussion} we discuss the implications of our findings. We summarize our results in Section~\ref{sec:Conclusions}.

\section{Data Reduction} \label{sec:methods}

Observations of Mrk~876 were taken from 2016 March through 2019 May. The observatories involved with this project are the Dan Zowada Memorial Observatory (Zowada), the Liverpool Telescope (LT), and the Las Cumbres Observatory (LCO). All images were processed using the standard pipelines for each observatory, which includes bias/dark subtraction and flat fielding. Each observatory uses the SDSS $ugri$ filters. Zowada and LCO both have the Pan-STARRS $z_s$ filter, while LT uses an SDSS $z$ filter. Between 2 to 5 images per filter were taken each night, depending on the instrument, filter, and weather conditions. 

Zowada is a robotic 0.5-meter telescope located outside of Rodeo, New Mexico \citep{2022arXiv220313400C}. It is owned and operated by Wayne State University. Located in the Canary Islands, LT is a robotic 2-meter telescope operated by the Astrophysics Research Institute of Liverpool John Moores University \citep{2004SPIE.5489..679S}. LCO is a global network of robotically operated telescopes \citep{2013PASP..125.1031B}. For this project, observations were taken from the 2-meter telescope at Haleakala Observatory (ogg02), as well as the two 1-meter telescopes (elp08 and elp06) at McDonald Observatory. Monitoring of Mrk~876 continued past 2019 May as part of a larger coordinated reverberation mapping campaign, and therefore those data are not used here for time lag analysis, but are only used to improve intercalibration between the light curves from the different telescopes. The LCO data from 2016-2018 was also used in \cite{Landt2022} to analyze the infrared reverberation signal. A summary of the observatories and observation epochs can be found in Table~\ref{table:epochs}. 

\begin{deluxetable*}{lccccc}
\tablecolumns{4}
\tablecaption{Summary of Observations
\label{table:epochs}}
\tablehead{
\colhead{Telescope} & \colhead{Epochs} & \colhead{Start Date} & \colhead{End Date} &
\colhead{Period Length} & \colhead{Cadence}
}
\startdata
 Zowada & 57 & 2019-01-24 & 2019-05-31 & 127 & 2.49\\ 
 LT & 175 & 2016-07-09 & 2018-08-12 & 764 & 3.53\\ 
 ogg02 (LCO) & 63 & 2016-02-15 & 2016-08-26 & 193 & 2.60\\ 
 elp08 (LCO) & 145 & 2016-03-31 & 2018-10-07 & 920 & 4.03\\ 
 Total & 383 & 2016-02-15 & 2019-05-31 & 1201 &2.53\\ 
\enddata
\tablecomments{An epoch is defined as a night on which observations were obtained. Each epoch may not have data from every filter, but indicates that at least one observation occurred. Period Length is the total number of days that each telescope's observation campaign lasted. Cadence is the average cadence of observations from the $g$ band in days excluding observational gaps of 30+ days.}
\end{deluxetable*}
The light curves are obtained using differential photometry. A selection of comparison stars are chosen to be contrasted with the brightness of Mrk~876. We assume that the combined flux from the comparison stars is constant over time.
We create an AGN light curve by calculating the flux of the AGN relative to the total flux of the comparison stars, allowing us to create the light curve for the AGN. Comparison stars are typically chosen such that they are 2-5 times brighter than the AGN in order to maximize the signal-to-noise ratio. All of the stars must be present in each telescope's field of view, meaning the choice of comparison stars was limited. Three stars were used for differential photometry based on the above factors. Different comparison stars are chosen for the $u$ band, since most stars are typically fainter in this band. 

\begin{figure*}
\centering
\includegraphics[width=1\textwidth]{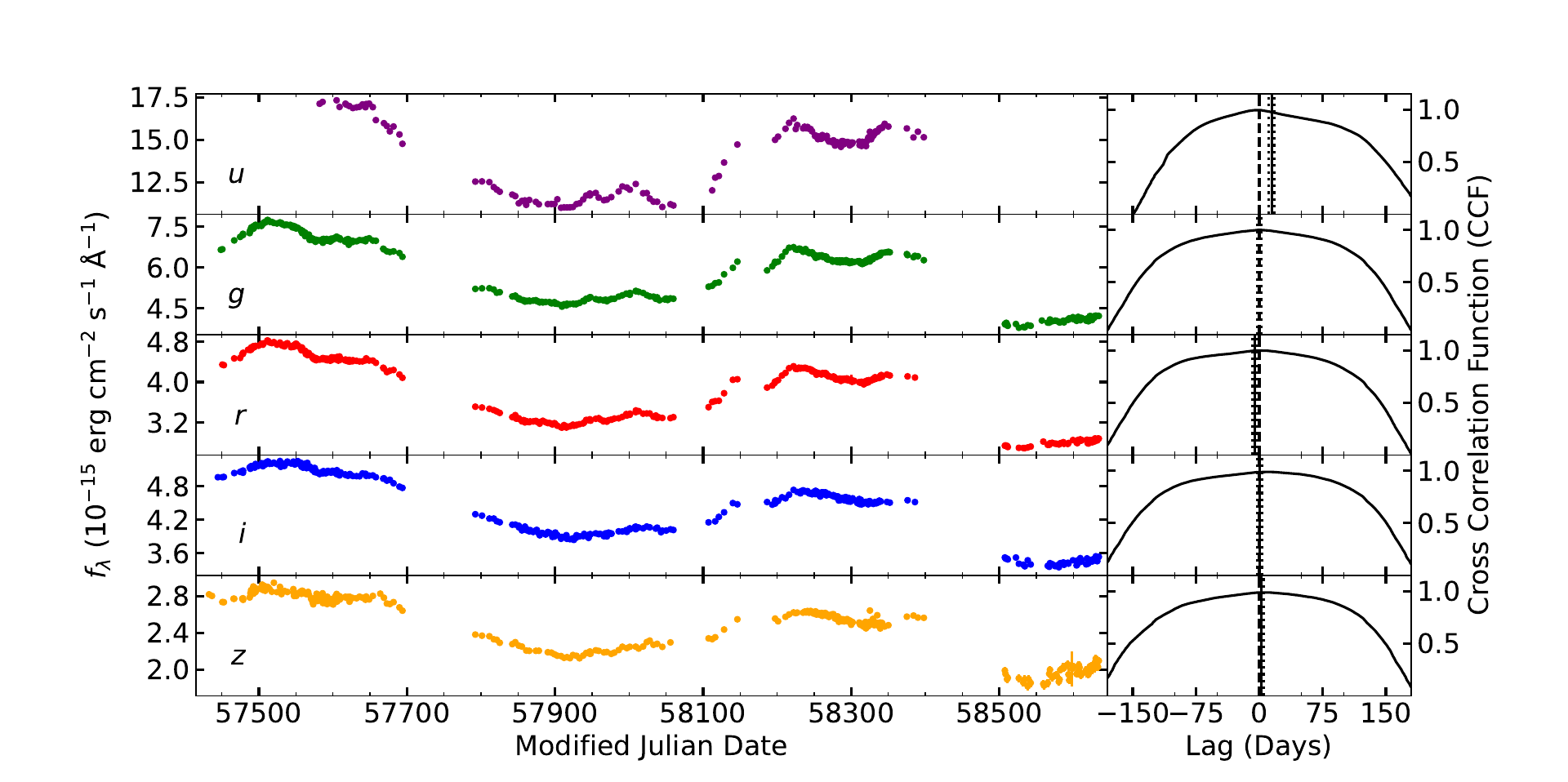}
\caption{The combined light curves of Mrk~876. The CCF is found for each light curve, measured with respect to the $g$ band. The dashed line is at 0~days, the solid and dotted lines show the PyCCF lag and corresponding uncertainties.}
\label{fig:light curve}
\end{figure*}

\begin{figure*}
\centering
\includegraphics[width=\textwidth]{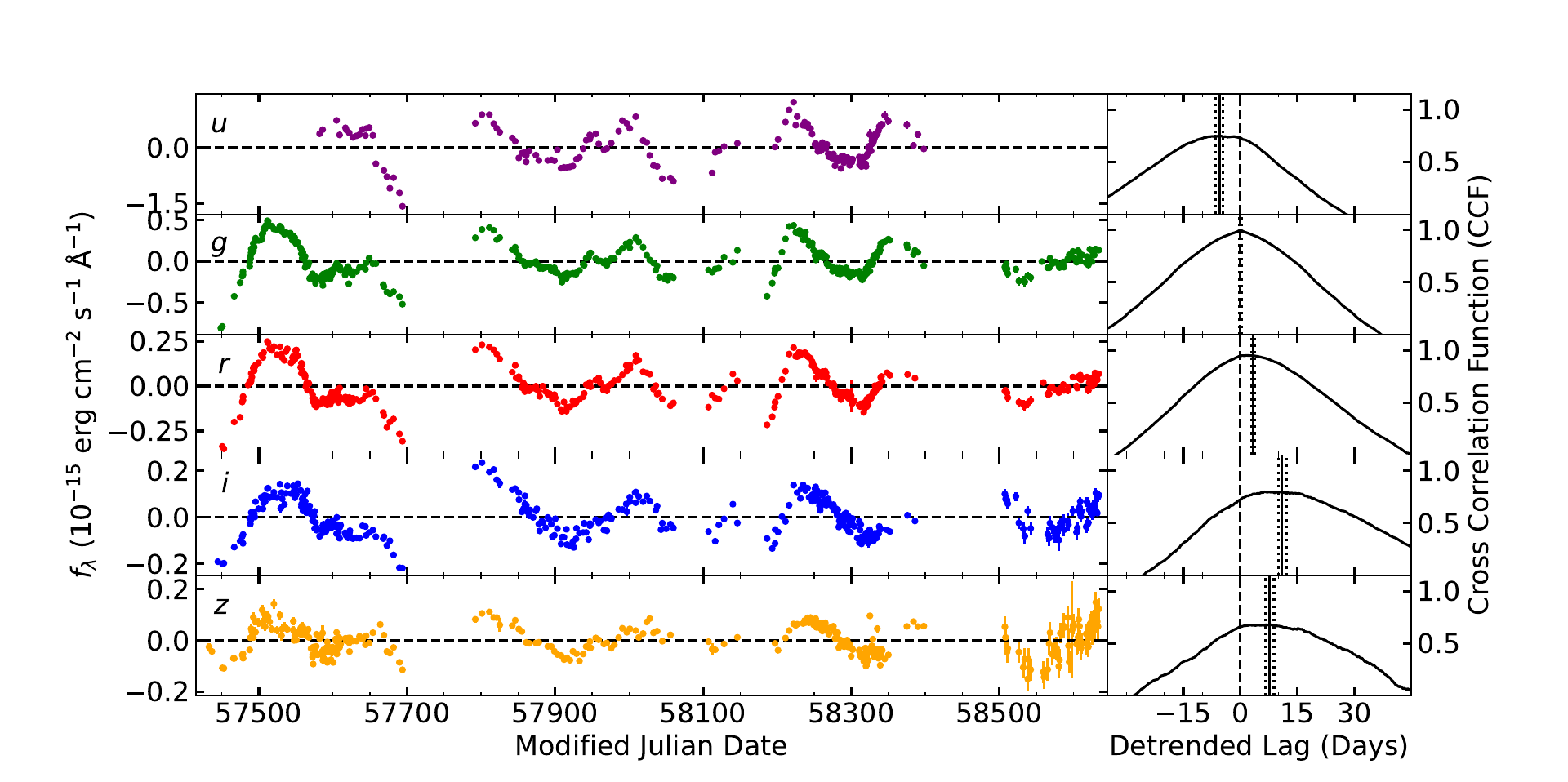}
\caption{The combined light curves of Mrk~876, detrended by subtracting a moving boxcar average of 100 days. The CCF is displayed following the same format as Fig. \ref{fig:light curve}}.
\label{fig:light curve_detrended}
\end{figure*}

For each image the stars are found using the photutils \citep{2021zndo...5796924B} module DAOStarFinder. Once found, we identify the comparison stars and AGN from the list of detected objects. A circular annulus and aperture are created for each object. The sizes of the apertures and annuli vary depending on the telescope. The annuli are chosen to have an inner radius of 20~pixels and an outer radius of 30~pixels for Zowada images. LCO and Liverpool have these annuli adjusted to match the same angular size based on their respective pixel scales. To choose the aperture size, we measure the signal-to-noise ratio for different aperture sizes and the fractional standard deviation of the comparison stars. We combine this in quadrature with the statistical uncertainty in flux. The aperture size with the lowest total flux uncertainty is chosen for each band and averaged to be used for each observatory. For Zowada, this is 5~pixels, for LT this is 8~pixels, for the LCO 1-meter observatories (elp06 and elp08) this is 11~pixels, and for the LCO 2-meter observatory (ogg02) this is 9~pixels. The median background is measured within the annulus and is scaled to the area within the aperture and subtracted. The average count rate is calculated from all of the observations of a specific filter taken on a given night, and these average observations are collected into a light curve for each band and each telescope. 

We combine and intercalibrate the light curves from all the telescopes using CALI \citep{2014ApJ...786L...6L}. CALI assumes the AGN variability is described by a dampened random walk process in order to interpolate between gaps in data and align multiple telescopes' data to a common scale. It applies both additive and multiplicative factors to the data to achieve this. A Bayesian framework with a diffusive nested sampling algorithm is used to determine the intercalibration factors. Additional systematic errors may exist, so CALI increases the uncertainty on all measurements with a systematic error term that is added in quadrature to the original uncertainties. The complete set of light curves can be found in the left panel of Fig.~\ref{fig:light curve}. When combined, we get an average cadence of 3.31~days in the $g$ band. When we exclude large observational gaps of $>$30~days, we have an average cadence of 2.53~days.

\begin{figure*}
\centering
\includegraphics[width=\textwidth]{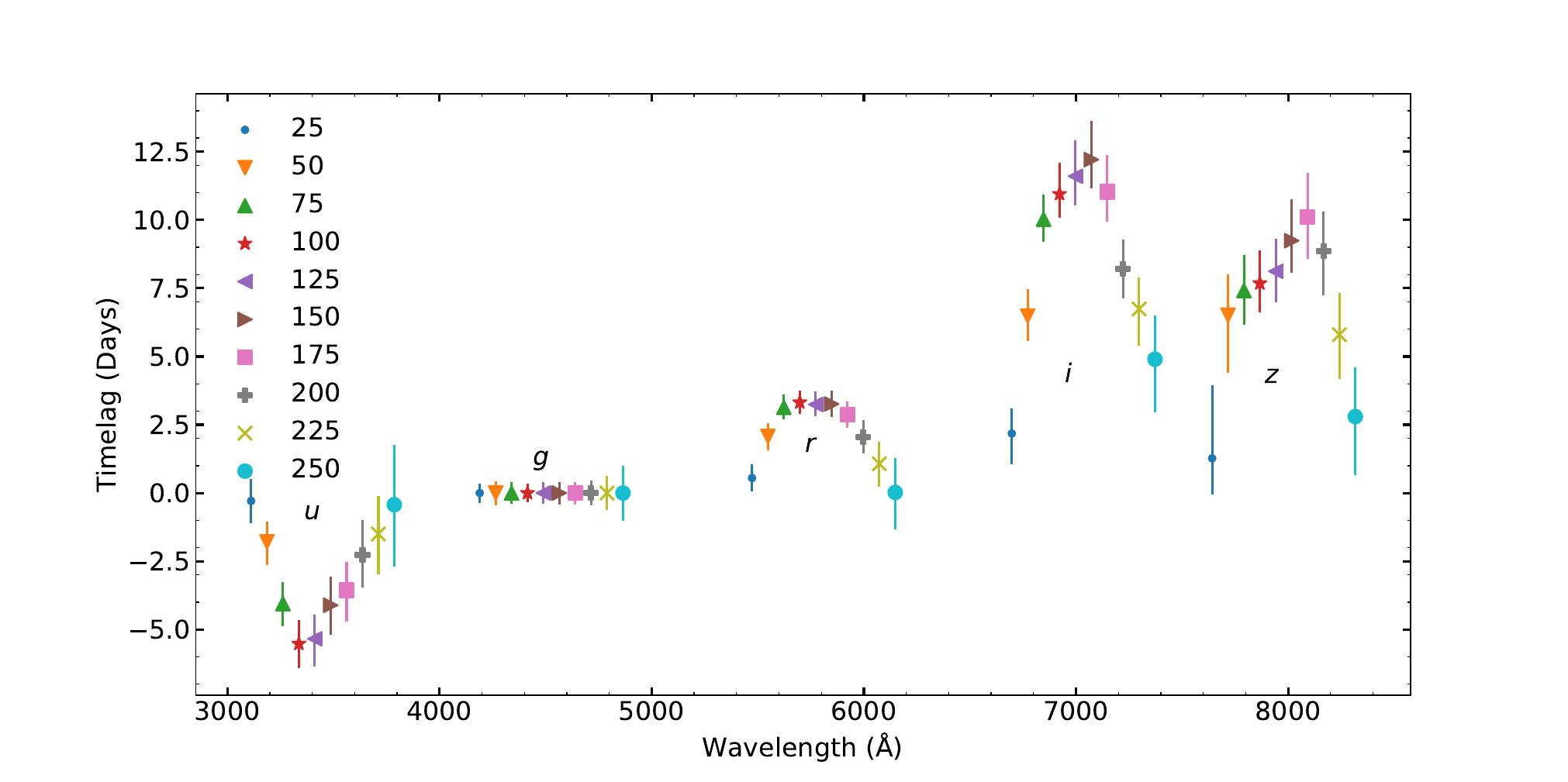}
\caption{A comparison between the different boxcar detrending widths used and the resultant timelags found using PyCCF. A detrending width of 100 days was chosen due to having the smallest overall uncertainties, and is shown here as the points with green stars. Widths from 75 to 175 agree to the lags found within uncertainty for all bands. The lags with uncertainties can be found in Table \ref{table:detrended_lag_comp}. All points have been fitted using the standard $ugriz$ bands, but have been spatially adjusted on this figure for clarity.}
\label{fig:DetrendTimeLagComp}
\end{figure*}

\begin{deluxetable*}{lccccc}
\tablecolumns{6}
\tablecaption{Time Lag Comparison
\label{table:detrended_lag_comp}}
\tablehead{
\colhead{Boxcar Width} & \multicolumn{5}{c}{Lag (days)} \\
(Days) & $u$ & $g$ & $r$ & $i$ & $z$ }
\startdata
0 (Base) & $14.60^{+3.72}_{-3.63}$ &  
$0.00^{+2.50}_{-2.52}$ &
$-5.17^{+2.85}_{-2.68}$ &
$0.43^{+2.87}_{-2.52}$ &
$2.31^{+3.10}_{-2.97}$ \\ 
25 & $-0.29^{+0.81}_{-0.80}$ & $0.00^{+0.35}_{-0.36}$ & $0.55^{+0.49}_{-0.48}$ & $2.18^{+0.92}_{-1.12}$ & $1.27^{+2.69}_{-1.31}$ \\
50 & $-1.79^{+0.73}_{-0.85}$ & $0.00^{+0.43}_{-0.44}$ & $2.07^{+0.48}_{-0.51}$ & $6.48^{+0.98}_{-0.90}$ & $6.51^{+1.49}_{-2.11}$ \\
75 & $-4.05^{+0.79}_{-0.83}$ & $0.00^{+0.39}_{-0.38}$ & $3.14^{+0.47}_{-0.45}$ & $10.03^{+0.89}_{-0.84}$ & $7.42^{+1.31}_{-1.27}$ \\
100 & $-5.52^{+0.87}_{-0.89}$ & $0.00^{+0.34}_{-0.34}$ & $3.32^{+0.42}_{-0.43}$ & $10.95^{+1.14}_{-0.88}$ & $7.68^{+1.20}_{-1.06}$ \\
125 & $-5.34^{+0.90}_{-1.01}$ & $0.00^{+0.39}_{-0.38}$ & $3.24^{+0.48}_{-0.44}$ & $11.6^{+1.33}_{-1.06}$ & $8.12^{+1.19}_{-1.12}$ \\
150 & $-4.11^{+1.05}_{-1.08}$ & $0.00^{+0.41}_{-0.41}$ & $3.26^{+0.49}_{-0.48}$ & $12.21^{+1.41}_{-1.05}$ & $9.24^{+1.51}_{-1.18}$ \\
175 & $-3.56^{+1.05}_{-1.16}$ & $0.00^{+0.40}_{-0.41}$ & $2.86^{+0.48}_{-0.47}$ & $11.03^{+1.34}_{-1.09}$ & $10.12^{+1.62}_{-1.55}$ \\
200 & $-2.26^{+1.28}_{-1.20}$ & $0.00^{+0.45}_{-0.46}$ & $2.04^{+0.62}_{-0.59}$ & $8.21^{+1.08}_{-1.07}$ & $8.87^{+1.43}_{-1.63}$ \\
225 & $-1.50^{+1.40}_{-1.48}$ & $0.00^{+0.63}_{-0.62}$ & $1.07^{+0.80}_{-0.83}$ & $6.74^{+1.15}_{-1.34}$ & $5.8^{+1.53}_{-1.63}$ \\
250 & $-0.43^{+2.20}_{-2.27}$ & $0.00^{+1.01}_{-1.01}$ & $0.02^{+1.25}_{-1.35}$ & $4.90^{+1.6}_{-1.94}$ & $2.80^{+1.81}_{-2.14}$ \\
\enddata
\tablecomments{A comparison of the PyCCF timelags for different detrending lengths. The first row with a detrend length of 0 days represents non-detrended data, which we will refer to as the base data in further analyses.}
\end{deluxetable*}

\section{Analysis and Results} 
\label{sec:results}

\subsection{Measured Time Lags}
\label{sec:timelaganalysis}
\begin{figure*}
\includegraphics[width=0.99\textwidth]{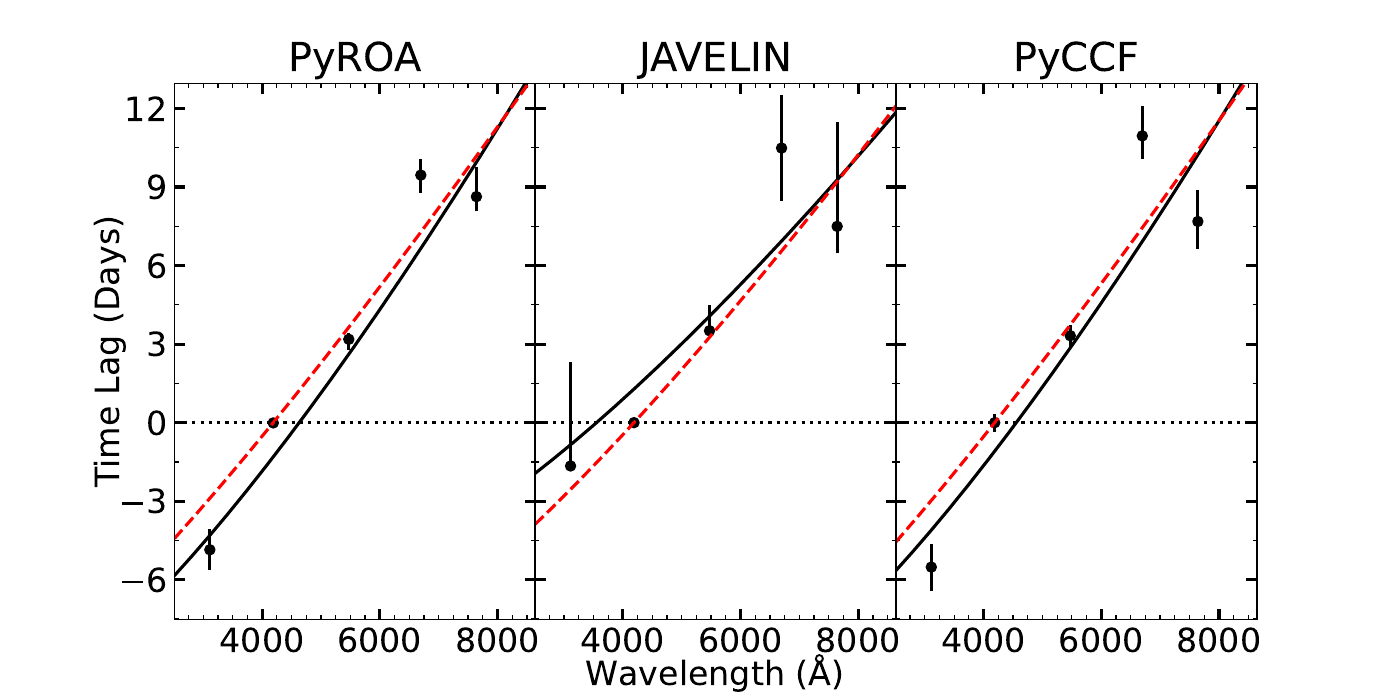}
\centering 
\caption{A comparison of the time lags found using PyROA, JAVELIN and PyCCF. The wavelengths have been redshift corrected. The solid black line is the best-fitting $\tau = \tau_0[(\lambda/\lambda_0)^\beta - y_0]$, with $\beta$ fixed to 4/3 while $\tau_0$ and $y_0$ are free. The red dashed line represents the $a$ = 0.998 analytic prescription for time lags as described by \cite{2021ApJ...907...20K}. This fit allows only the X-ray corona height ($H$) to be a free parameter, while black hole mass, X-ray luminosity, and mass accretion rate ($\dot{m}$) are fixed. When ($\dot{m}$) is allowed to be a free parameter, the fit becomes unconstrained for both spin cases. For more details, see Sec. \ref{sec:TheoreticalTimeLags}.}
\label{fig:lags}
\end{figure*}

The light curves for the $ugriz$ filters can be found in Fig.~\ref{fig:light curve}. To quantify the variability, the excess variance $F_{\rm var}$ \citep{2002ApJ...568..610E, 2003MNRAS.345.1271V} is calculated for each of the light curves. Strong variability is observed, with a variability amplitude of 13\%, 19\%, 16\%, 13\%, and 12\% in the $u$, $g$, $r$, $i$, and $z$ bands respectively.
To determine the time lags, the cross correlation function (CCF) is found for these light curves using the $g$ band. This band is chosen as the comparison band because it has the best-sampled light curve and has the highest variability amplitude. The CCF is found using the Python module PyCCF \citep{2018ascl.soft05032S}, which follows the \cite{2004ApJ...613..682P} implementation of the ICCF technique to determine lag uncertainties. PyCCF calculates the CCF of two unevenly sampled light curves using interpolation to fill in the gaps between data points. The mean of the CCF at 80\% of the CCF's peak (the centroid) is taken as the time lag between the two respective bands. We find in Fig.~\ref{fig:light curve} that the CCF of Mrk~876 is extremely broad, with a near-flat top spanning several hundred~days. This indicates the lag is dominated by long-term variations observed in the light curve. The broad CCF prevents a robust reverberation lag measurement.
The data are detrended by subtracting a moving boxcar average to remove this long-term trend. The process of detrending AGN for improved short-term lag measurements is a common practice \citep[e.g.,][]{1999PASP..111.1347W, 2004ApJ...613..682P, 2014MNRAS.444.1469M, 2015ApJ...806..129E, 2016ApJ...821...56F, 2020MNRAS.494.4057P, 2020MNRAS.498.5399H}. The data were boxcar subtracted with a variety of boxcar widths ranging from 25-250~days, in intervals of 25~days. The lags of each set of detrended light curves were calculated using PyCCF as well. All boxcar widths between 75 and 175 days produced CCF lags consistent within 1$\sigma$ uncertainties. The boxcar width with the lowest average lag uncertainty was 100~days, and therefore was chosen for continued analysis. This process does not guarantee that the resulting time lags are true representations of the intrinsic lag between the different wavelengths. While the majority of boxcar lengths tested produce time lags that agree within 1$\sigma$, other detrending approaches may produce different results.
The light curves and CCF of the 100~days detrended data can be found in Fig.~\ref{fig:light curve_detrended}. A plot comparing all of the recovered time lags versus detrend lengths is shown in Fig.~\ref{fig:DetrendTimeLagComp}, with the lags given in Table~\ref{table:detrended_lag_comp}. The $g$ lags are used for time-lag comparison against the other bands. We also measure the lag of the $g$ band against itself, allowing the $g$ band to have a lag beyond exactly 0~days. We use the detrended light curves for most analyses going forward. When we use the non-detrended data, we will refer to it as the base data.
The light curves span several years, and due to Mrk~876's position in the sky there are annual seasonal gaps when Mrk~876 was not visible. The CCF method simply performs a linear interpolation between the gaps. However, more sophisticated methods, e.g. JAVELIN and PyROA, have been developed to use the variability properties of the light curves to inform a more realistic interpolation. JAVELIN \citep{2011ApJ...735...80Z, 2016ApJ...819..122Z} uses a dampened random walk model for the power spectrum of the light curves and assumes a top-hat transfer function. A recent comparison between CCF and JAVELIN methods for determining lags is presented by \cite{2020MNRAS.491.6045Y}. We use the Python 3 implementation of JAVELIN for our analysis. PyROA \citep{2021MNRAS.508.5449D} uses a running optimal average to determine a model for the driving light curve, and assumes that each light curve is a time-shifted and flux-scaled version of this model. The Bayesian Information Criterion is then used to determine how much smoothing is required for the data, and model parameters are estimated using Markov Chain Monte Carlo. The time lags calculated using each method are given in Table~\ref{table:lags}, while in Fig.~\ref{fig:lags} we show the lags as a function of wavelength. We also give the maximum correlation coefficient ($R_{\rm max}$) in Table~\ref{table:lags}. All the light curves are well correlated. The lags generally agree between each method within their errors.
\begin{deluxetable}{cccccc}
\tabletypesize{\footnotesize}
\tablewidth{0.99\textwidth}
\tablecolumns{6}
\tablecaption{Detrended Time Lags and Light Curve Properties
\label{table:lags}}
\tablehead{
\colhead{Method} &\colhead{$u$}&\colhead{$g$}&\colhead{$r$}&\colhead{$i$}&\colhead{$z$}}
\startdata
 R$_{\rm max}$ & 0.75 & 1.0 & 0.95 & 0.80 & 0.68\\
 PyCCF & $-5.52^{+0.87}_{-0.89}$ &$-0.0^{+0.34}_{-0.34}$ &$3.32^{+0.42}_{-0.43}$ &$10.95^{+1.14}_{-0.88}$ &$7.68^{+1.20}_{-1.06}$ \\
 JAVELIN & $-1.66^{+3.96}_{-0.06}$ &$-0.0^{+0.00}_{-0.00}$ &$3.51^{+0.99}_{-0.03}$ &$10.49^{+2.01}_{-2.03}$ &$7.5^{+3.97}_{-1.03}$ \\
 PyROA & $-4.86^{+0.78}_{-0.78}$ &$-0.01^{+0.16}_{-0.16}$ &$3.18^{+0.25}_{-0.40}$ &$9.46^{+0.62}_{-0.69}$ &$8.63^{+1.15}_{-0.56}$ \\
\enddata
 \tablecomments{R$_{\rm max}$ is the maximum value of the cross correlation coefficient between the two light curves (calculated with respect to $g$). A value of 1 is maximal correlation, while a value of 0 means no correlation between the signals. The lags for PyCCF, JAVELIN, and PyROA are all given in units of days and are calculated with respect to the $g$ band.}
\end{deluxetable}

\subsection{Theoretical Time Lags}
\label{sec:TheoreticalTimeLags}
The expected time lag $\tau$ can be calculated from a simple model of the accretion disk. Assuming a standard lamppost-like X-ray corona ionizing the accretion disk, the time lag should relate to the radius of the disk $R$ as $\tau \sim R/c$. For a geometrically thin, optically thick accretion disk \citep{1973A&A....24..337S}, the temperature profile follows $T(R) \propto (M\dot{M})^{1/4}R^{-3/4}$. Assuming blackbody radiation so that $\lambda \propto T^{-1}$, one finds a relationship between the time lag and the emitted wavelength of light:
\begin{equation}
\tau \propto (M\dot{M})^{1/3}T^{-4/3} \propto (M\dot{M})^{1/3}\lambda^{4/3}.
\label{eq:lageq}
\end{equation}
We test this relation by fitting the following function: 
\begin{equation}
\tau = \tau_0[(\lambda/\lambda_0)^\beta - y_0],
\label{eq:wavlag}
\end{equation}
where $\tau_0$ is the normalization parameter and is measured in days, $\beta$ is the relationship of wavelength to the measured timelags, $\lambda$ is the wavelength of the band being observed, and $\lambda_0$ is the reference wavelength band used, which for this study is $\lambda_0 = 4770$\AA\ (the effective wavelength of the $g$ band). The adjustment factor $y_0$ is present to prevent the lags from being exactly zero at $\lambda_0$, with its value normally being around 1. A standard thin disk predicts $\beta$ = 4/3. We fit the lag-wavelength relation both fixing $\beta$ = 4/3 and allowing it to be a free parameter in the fit. Best-fits are shown in Fig.~\ref{fig:lags}, and given in Table~\ref{table:alphas}. When $\beta$ is left as a free parameter it drops below 1, leading to an unreasonably large value for $\tau_0$. As such, it is not included in the figure. This equation holds the assumption that the x-ray emitting component's height is small relative to the radius of the accretion disk. If this is not true, then this acts as a lower limit to the measured lags.

We also fit to the lags the analytic prescription described by \cite{2021ApJ...907...20K}. This prescription predicts time lags using five different parameters. These are the black hole mass, mass accretion rate ($\dot{m}$), X-ray luminosity in the 2-10 keV range, X-ray corona height ($H$), and black hole spin ($a^*$). We fit our observed time lags for when the spin parameter $a^*$ = 0.998 and $a^*$ = 0. The black hole mass is estimated to be $2.2 \times10^8$~M$_{\odot}$ \citep{2004ApJ...613..682P, 2015PASP..127...67B}, and the X-ray luminosity in the 2-10 keV range is 10$^{44.11}$ erg/s from archival XMM-Newton data \citep{2018ApJ...868...10L}. We estimate the mass accretion rate using our calculation of the Eddington accretion rate ($\dot{m}_E$), which is described in Sec. \ref{sec:normalization} as a part of the analysis on the normalization parameter $\tau_0$. The initial fits are done fixing $\dot{m}$ to be this value while $H$ remains the only free parameter. This fit fails for $a^*$ = 0, with the corona height $H$ becoming non-physical. This agrees with previous investigations indicating that Mrk 876's SMBH has a high spin parameter \citep{bottacini2022}. We also perform the fitting allowing $\dot{m}_E$ to be a free parameter. For this regime we used a grid search to find the lowest possible $\chi^2$ value. We search from 0 -- 250~$R_G$ for $H$ and between 0 -- 0.75 Eddington accretion for $\dot{m}_E$. The upper and lower bounds for the uncertainty of the measurement are determined by finding the value of the parameters when $\chi^2$ is 2.3 above the lowest value found. When left as free parameters, uncertainties on $\dot{m}_E$ and H are largely unconstrained, but agree within 1-sigma (for both zero spin and maximally spinning cases). We therefore only give the parameters when $\dot{m}_E$ is fixed and $a^*$~=~0.998 in Table \ref{table:alphas}.


\begin{deluxetable}{cccc}
\tabletypesize{\footnotesize}
\tablewidth{0.99\textwidth}
\tablecaption{Fitted Accretion Disk Properties
\label{table:alphas}}

\tablehead{
\colhead{Method} &\colhead{$\tau_0$ (days)} &\colhead{ $H$ ($R_G$)}&\colhead{$\dot{m}_E$}
}
\startdata
PyCCF & 9.22 $\pm$ 2.64 & 54.0 $\pm$ 56.0 &  0.416\\ 
JAVELIN & 6.50 $\pm$ 2.00 & 28.0 $\pm$ 55.0 &  0.416\\  
PyROA & 9.16 $\pm$ 1.38 & 48.0 $\pm$ 25.0 &  0.416\\ 
\enddata
\tablecomments{Comparison of the fitted parameters between PyCCF, JAVELIN, and PyROA. The normalization parameter $\tau_0$ from Eqn.~\ref{eq:lageq} is found when $\beta$ is fixed to be 4/3, as expected from thin disk theory. These are plotted as the solid black lines in Fig.~\ref{fig:lags}. The X-ray corona height ($H$) is found when the mass accretion fraction ($\dot{m}$) is fixed to be 0.416 for a spin parameter of $a^*$ = 0.998. For more details, see Sec.~\ref{sec:TheoreticalTimeLags}.}
\end{deluxetable}
\subsection{Calculation of Normalization Parameter $\tau_0$}\label{sec:normalization}
One noted problem among AGN reverberation mapping campaigns is the measured value of normalization parameter $\tau_0$. We can estimate the expected value of $\tau_0$ for the thin disk model using estimates of the black hole mass and mass accretion, and compare to the measured value \citep{2016ApJ...821...56F}. The majority of campaigns have found that the fitted value of $\tau_0$ is 2-3 times larger than the expected/calculated value \citep[e.g.,][]{2014ApJ...788...48S, 2015ApJ...806..129E, 2016ApJ...821...56F, 2017ApJ...836..186J}. This implies that the accretion disk itself is 2-3 times larger than the standard thin disk model predicts. Alternatively, some other aspect of the AGN system is interfering with measurements and creating a falsely large measurement of $\tau_0$. 

\cite{2016ApJ...821...56F} parameterize the following equation for calculating $\tau_0$:
\begin{equation}
\tau_0 = \frac{1}{c} \left(X\frac{k \lambda_0}{hc}\right)^{4/3} \left[\left(\frac{GM}{8 \pi \sigma}\right) \left(\frac{L_{\mathrm{Edd}}}{\eta c^2}\right) \left( 3+\kappa \right) \dot{m}_{\rm E} \right]^{1/3}.
\label{eq:alpha}
\end{equation}
To calculate this value for Mrk~876, several assumptions are made. It is assumed that the X-rays and viscous heating contribute roughly the same amount of energy to the disk, so that the radiative efficiency for converting rest mass into radiation $\eta$ = 0.1 and the local ratio of external heating to internal heating $\kappa$= 1. The factor $X$ is a multiplicative factor of order unity, and is determined from temperature $T$ and the wavelength measured via $T = Xhc/k\lambda$. This factor helps account for how temperature relates to the wavelength emitted for a given radius, and is influenced by the choice of geometry in the disk. For a flux-weighted mean radius, $X = 2.49$ \citep{2016ApJ...821...56F}, but taking variation of the disk emission into account leads to $X = 5.04$ \citep{2018MNRAS.473...80T}. 

To calculate the Eddington accretion rate ($\dot{m}_E$), one must estimate the bolometric luminosity. We use the standard bolometric correction via $L_{\rm bol} \sim 9\lambda L_{\lambda}(5100\angstrom)$ \citep{2000ApJ...533..631K}. However, since 5100$\angstrom$ is not a central wavelength of the SDSS filters, the $g$ band is used as the nearest available approximation. The $g$-band data are converted from relative flux to magnitudes using the comparison stars, which have magnitudes from the APASS catalogue. The AGN magnitudes are then converted into fluxes and extinction corrected with an $E(B-V)$ of 0.027 \citep{2007ApJS..170...33P} using Cardelli's extinction law \citep{1989ApJ...345..245C}, and corrected to rest-frame fluxes using $z = 0.1385$ \citep{2011MNRAS.416.2840L}. We assume a luminosity distance of $D_{\rm L} = 588.4$~Mpc to calculate $L({5100\angstrom})$ from the dereddened, rest-frame flux. The Eddington luminosity is calculated assuming a black hole mass of $2.18\times10^8$~M$_{\odot}$ \citep{2015PASP..127...67B}. For our bolometric luminosity, $L_{\rm bol} = 1.144 \times 10^{46}$~erg~s$^{-1}$, we determine an Eddington fraction of $\dot{m}_{\rm E} = 0.416$. Substituting into Eqn.~\ref{eq:alpha} we calculate $\tau_0 = 2.57$~days and $6.58$~days for $X = 2.49$ and $X = 5.04$ respectively.



 \subsection{Spectral Analysis}\label{sec:spectralanalysis}
\begin{figure*}
\includegraphics[width=0.9\textwidth]{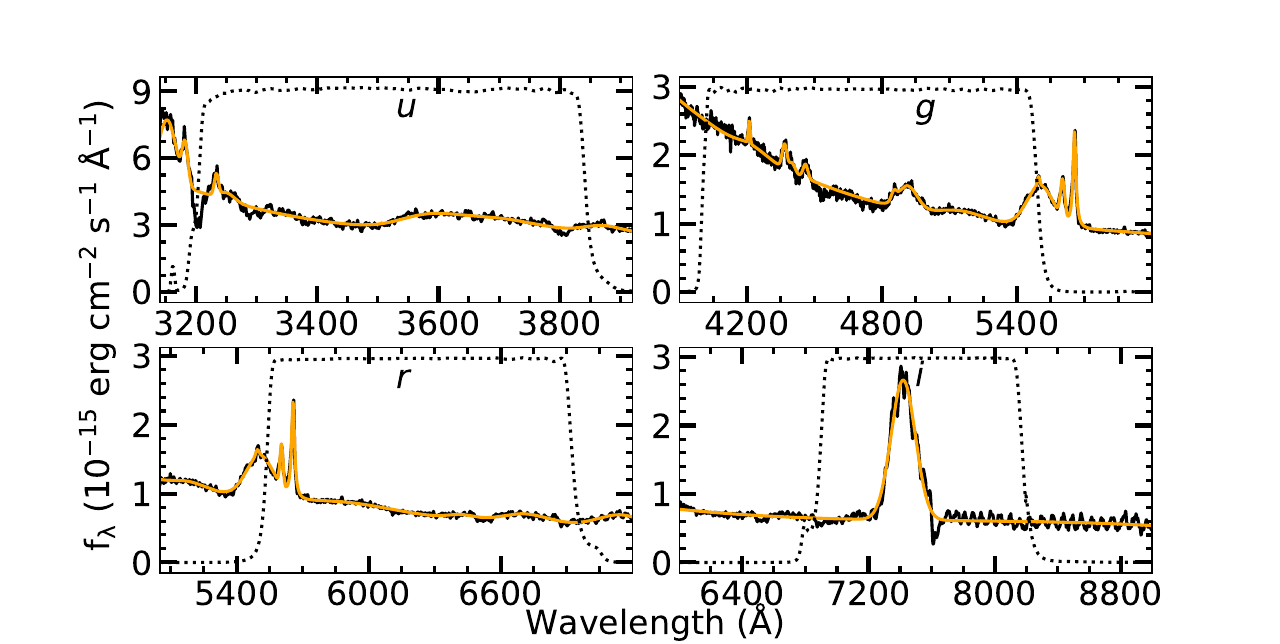}
\centering 
\caption{Spectra taken on 2019-07-02 of Mrk~876. The data are shown in black while the fitted model, containing both a continuum and multiple Gaussian and Lorentzian components, is in orange. The dotted line is the throughput of the $u$, $g$, $r$, and $i$ bands. We only find significant emission line contribution from the H$\alpha$ line present in the $i$ band. The exact contribution from the continuum and emission lines can be found in Table \ref{table:throuput}.}
\label{fig:spectra_allfilters}
\end{figure*}

 We note that in the time lag analysis the $i$-band lags (Fig.~\ref{fig:lags}) are consistently offset from the rest of the trend. At the redshift of Mrk~876, H$\alpha$ is close to the effective wavelength of the $i$ band. Other emission lines may be affecting the other lags as well. Fig.~\ref{fig:spectra_allfilters} shows spectra taken on 2019-07-02 (just after the end of the campaign) from the LCO Haleakala Observatory (FTN) overlaid with the filters used to take the data. The $z$ band shows no significant emission line contribution. The $u$, $g$, and $r$ bands show some emission lines. To determine total flux contribution, we modeled the emission lines using astropy Spectrum1D models. These are shown in Fig.~\ref{fig:spectra_allfilters} as the orange lines. The total contribution is summed up from each model for each point, using the throughput of each filter as a modifier of the total strength of the emission. The percentage of flux that comes from the continuum versus the emission lines are found in Table \ref{table:throuput}. Note that we only factor in the broad emission lines to the emission line percentage, and not any potential contribution from the diffuse continuum which is also thought to originate from the BLR. We find that the majority of filters see a small amount of emission line contribution, but not enough to warrant additional consideration. The exception to this is the $i$ band, where we find that H$\alpha$ contributes 33\% of the total flux. To ensure that the presence of H$\alpha$ was consistent throughout the campaign, we also analyze spectra taken from the start of the campaign in 2016. We find the H$\alpha$ line contributes 29\% of the total flux in 2016, confirming that H$\alpha$ is a strong, consistent presence throughout the entire monitoring campaign. 
 

\begin{deluxetable}{ccc}
\tabletypesize{\normalsize}
\tablewidth{1.2\textwidth}
\tablecolumns{2}
\tablecaption{Emission Line Contributions by Filter
\label{table:throuput}}
\tablehead{
\colhead{Filter} &\colhead{Continuum} &\colhead{Emission Line}}
\startdata
 $u$ & 91\% & 9\% \\ 
 $g$ & 94\% & 6\% \\ 
 $r$ & 85\% & 15\% \\
 $i$ & 67\% & 33\% \\
\enddata
\tablecomments{The percentage of continuum and emission line contribution to the overall image from the modeled LCO spectra found in Fig.~\ref{fig:spectra_allfilters}.}
\end{deluxetable}


 However, a 30\% contribution by H$\alpha$ does not imply a 30\% increase in lag. It is a common expectation that to zeroth order, the continuum lag and the H$\alpha$ lag will combine weighted by their flux, however, simulations have shown that F$_{\mathrm{var}}$ is the dominating factor \citep{2016ApJ...821...56F}. The H$\alpha$ lag for Mrk~876 has been measured to be $43^{+40}_{-22}$~days, with a measured variability amplitude smaller than the continuum variability \citep{2000ApJ...533..631K}. However, it is more complicated than this, as the lag also depends on the variable flux, properties of the driving light curve, and the shapes of the transfer functions. Detailed simulations would be needed to properly assess this and are beyond the scope of this paper. Given that the H$\alpha$ flux is a significant fraction of the flux in the $i$ band, it is plausible to attribute the excess $i$-band lag to the H$\alpha$ line.

\subsection{Flux-Flux Analysis}
\label{sec:fluxfluxanalysis}
\begin{figure*}
\includegraphics[width=\textwidth]{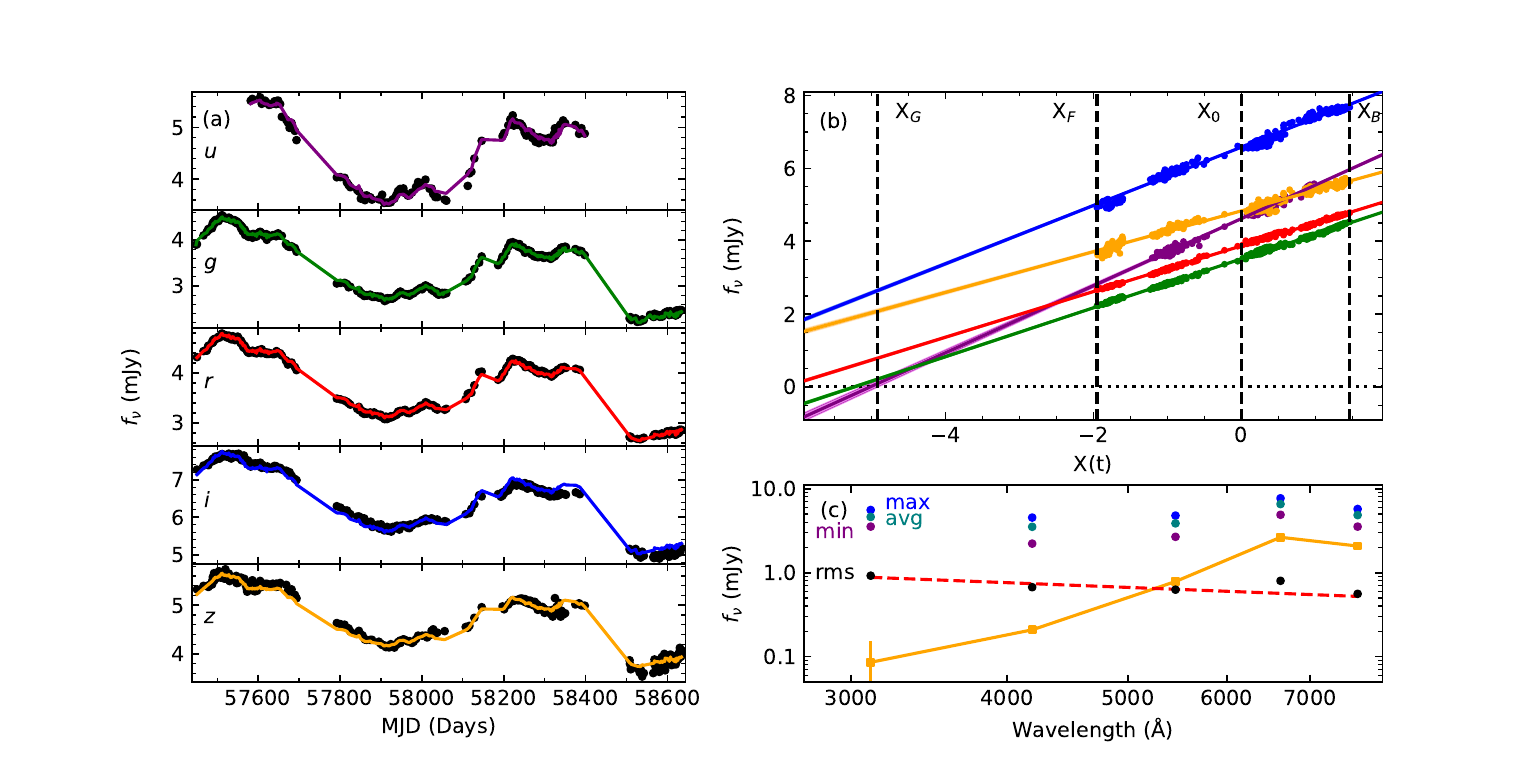}
\centering
\caption{Flux-Flux analysis on the base dataset of Mrk~876. (a) Redshift and Galactic extinction corrected fluxes in mJy for each filter. Data points are black, while the colored lines are the model (as described by Eqn.~\ref{eq:fluxflux}) overlaid. (b) The flux-flux analysis, with each band having the same colors as shown in (a). $X_G$ is where the first band's lower uncertainty crosses the $x$-axis, which in this case is the $u$ band. This is done so that a value for the constant component can be measured for this band.} (c) Maximum (in blue), average (in teal), and minimum (in purple) flux values for each filter. Rms values calculated from Eqn.~\ref{eq:fluxflux} are given in black. The red dashed line is a $f_\nu \propto \lambda^{-1/3}$ fit to the rms values. The solid orange line labeled const. is the constant contribution for the fluxes, measured by where each filter's line crosses $X_G$. The values for each filter are listed in Table~\ref{table:fluxflux}.
\label{fig:fluxflux_0.027}
\end{figure*}

To determine the spectral energy distribution (SED) of the variable flux, we perform a flux-flux analysis on the base data similar to \cite{2020ApJ...896....1C} \citep[see also][]{2017ApJ...835...65S, 2020MNRAS.498.5399H}. The photometric light curves are first flux-calibrated using the magnitudes of the comparison stars as found in the APASS catalogue \citep{2018AAS...23222306H} for the $g$, $r$, and $i$ bands, and the Pan-STARRs catalogue \citep{2018AAS...23110201C} for the $z$ band. Neither catalogue contained our $u$-band comparison stars, so as a proxy we use an observation from the Neil Gehrels Swift Observatory \citep{2004ApJ...611.1005G} of Mrk~876. The flux-calibrated light curves are corrected for Galactic absorption with an $E(B-V)$ of 0.027 \citep{2007ApJS..170...33P}. We use the extinction law of \cite{1989ApJ...345..245C}, and adjust the data to rest-frame flux. We then perform the flux-flux analysis by breaking the flux into constant and variable components, representing the galaxy and the AGN respectively, using the following formula:
\begin{equation}
f_{\nu}(\lambda, t) = A_{\nu}(\lambda) + R_{\nu}(\lambda) X(t).
\label{eq:fluxflux}
\end{equation}
$A_{\nu}$ is the average spectrum, $R_{\nu}$ is the rms spectrum, and $X(t)$ is a dimensionless light curve normalized to a mean of 0 and a standard deviation of 1. The light curves and fits for the non-detrended data are shown in Fig.~\ref{fig:fluxflux_0.027}, Panel (a). The flux-flux relations are shown in Panel (b). To estimate the galaxy contribution to the different bands, we extrapolate the fits to where the uncertainty envelope of the shortest wavelength band crosses $f_{\nu} = 0$ which we define as $X(t) = X_G$. This serves as a reference point for the other bands, and determining $f_{\nu}$ at $X(t) = X_G$ provides a lower limit on the constant component in each band. The dashed lines $X_F$ and $X_B$ represent the lowest and highest points found from all filters, and $X_0$ is given as reference. Panel~(c) of Fig.~\ref{fig:fluxflux_0.027} shows the maximum, minimum and average SED of Mrk~876 during the monitoring, along with the variable (rms) and constant spectral components determined from the flux-flux analysis. Table~\ref{table:fluxflux} gives the values determined with the flux-flux analysis. The rms spectrum is consistent with $f_\nu \propto \lambda^{-1/3}$ expected for an accretion disk spectrum. An excess in the variable spectrum in the $i$ band is seen, which would be consistent with a significant variable broad H$\alpha$ line contributing in that band. There is also an additional constant component in the $i$ band shown in Panel (c), indicating there is another source of continuum emission beyond the accretion disk.

\begin{deluxetable*}{cccccc}
\tabletypesize{\footnotesize}
\tablewidth{0.99\textwidth}
\tablecolumns{6}
\tablecaption{Flux-Flux Results
\label{table:fluxflux}}
\tablehead{
\colhead{} &\colhead{$u$}&\colhead{$g$}&\colhead{$r$}&\colhead{$i$}&\colhead{$z$}}
\startdata
Max & $5.58 \pm 0.02$ & $4.53 \pm 0.01$ & $4.79 \pm 0.01$ & $7.69 \pm 0.02$ & $5.74 \pm 0.04$ \\
Mean & $4.61 \pm 0.00$ & $3.52 \pm 0.00$ & $3.87 \pm 0.00$ & $6.56 \pm 0.0$ & $4.85 \pm 0.00$ \\ 
Min & $3.54 \pm 0.02$ & $2.22 \pm 0.03$ & $2.68 \pm 0.03$ & $4.9 \pm 0.07$ & $3.54 \pm 0.09$ \\ 
Constant & $0.09 \pm 0.07$ & $0.21 \pm 0.01$ & $0.78 \pm 0.01$ & $2.64 \pm 0.02$ & $2.08 \pm 0.04$ \\
RMS & $0.92 \pm 0.02$ & $0.67 \pm 0.00$ & $0.63 \pm 0.00$ & $0.80 \pm 0.01$ & $0.56 \pm 0.01$ \\
\enddata
\tablecomments{Base light curve flux-flux analysis values, which are shown in Fig.~\ref{fig:fluxflux_0.027}, panel~c. All units are in milliJanskies (mJy).}
\end{deluxetable*}


 \section{Discussion} \label{sec:discussion}
 We performed photometric monitoring of the AGN Mrk~876 over a 3 year period, during which is exhibited large amplitude (12 -- 19\%) variability in each of the $ugriz$ bands (see light curves in Fig.~\ref{fig:light curve}). We looked for lags between the different bands, as would be expected from an accretion disk reverberation scenario where ionizing radiation drives variability at longer wavelengths, with the hottest, inner (ultraviolet) part of the disk responding before the cooler, outer (optical) region of the disk.
 
 Initial CCF lags recovered from PyCCF reveal a long-term variability ($>$100$\sim$days) that dominates over the short term variability expected from accretion disk reverberation. AGN are known to vary on longer time scales, and to accurately recover the reverberation lags expected on shorter timescales (days), this long-term variability needs to be removed. We remove the long-term variability by subtracting a moving boxcar average from the light curves. To do this, a moving boxcar average is subtracted from the data. We test a range of widths to see which provided the lowest uncertainties in the resulting time lags. We find that detrend widths of 75-175~days produce time lags that agree within 1$\sigma$ uncertainties, and select 100~days as this width provided the lowest uncertainties. The base data CCF distributions are shown to the right of the light curves in Fig.~\ref{fig:light curve}, and the detrended light curves and their CCFs can be found in Fig.~\ref{fig:light curve_detrended}. Once the long-term trends are removed, the resulting CCF is significantly narrower, allowing a precise recovery of the reverberation lags. The measured lags from all tested detrending lengths can be found in Table~\ref{table:detrended_lag_comp}. 
 
 The source of the long-term variability may be changes in the accretion rate. This acts on the viscous timescale, predicted to be on the scale of hundreds of days for the optical emitting region \citep{1991ApJ...371..541K}. As more reverberation mapping studies are undertaken, it is becoming apparent that the lamppost model does not adequately explain all variability seen in some AGN light curves. It is possible that many AGN exhibit these long-term trends, as many observation campaigns detect some kind of long-term variation often associated with accretion disk flow or broad-line region interference \citep{2008MNRAS.389.1479A, 2014MNRAS.444.1469M, 2009MNRAS.394..427B, 2020MNRAS.498.5399H}. It is difficult to perform the lengthy monitoring needed to capture these long-term variations with traditional observing on a single telescope. With the rise of robotic observatories like LT, LCO, and Zowada, as well as all-sky surveys, such as ASAS-SN, ATLAS, ZTF, PAN-STAARS, and CRTS, more studies like this one will be possible. 
 
 The wavelength dependence of the detrended lags are shown in Fig.~\ref{fig:lags}. As a comparison, the lags are also found with the more sophisticated JAVELIN and PyROA techniques. We fit the relation $\tau \propto \lambda^{\beta}$ to all sets of lags. The lags recovered by PyCCF, JAVELIN, and PyROA are generally in agreement within uncertainty, and all recovered lags are well represented by $\tau \propto \lambda^{4/3}$. They are consistent with the expected wavelength dependence for a Shakura-Sunyaev geometrically thin optically thick accretion disk with an illuminating central source. All methods find an excess in their $i$-band lags, deviating from an extrapolation of the trend in the other wavebands. The lags from different methods are given in Table~\ref{table:lags}.
 
 Notably, we do not detect a $u$-band excess lag in our detrended time lag measurements. These excesses have been detected in a number of reverberation mapping studies \citep{2015ApJ...806..129E, 2017ApJ...840...41E, 2016ApJ...821...56F, 2021iSci...24j2557C}, however they also have not been detected as well \citep{2021ApJ...922..151K}. Without UV/X-ray monitoring, we lack the wavelength coverage to determine if Mrk~876 truly lacks a $u$-band lag excess. However, if the source of the $u$-band excess is from the BLR \citep{2001ApJ...553..695K}, then it is possible that our detrending process has already removed this contribution. Looking at the base lightcurve timelags in Table~\ref{table:detrended_lag_comp}, we see that $u$-band lag has the largest lag of all the bands at 14.68~days. This could imply that the BLR $u$-band emission operates on timescales of around 100 days, and that its contribution is removed by our detrending process. However, the large flat-topped CCFs that are produced prevent a robust lag measurement, so this result should be taken with caution.
 
 The boundary of the accretion disk and the dusty torus can be examined using our data. We can extrapolate what the values for longer wavelength lags would be, assuming they follow the $\tau = \lambda^{4/3}$ trend predicted for a geometrically thin optically thick disk. We compare the values we predict against the values determined from fitting the near-infrared rms spectrum by \cite{Landt2022}. The implied radii are measured to be $\sim$ 25~light-days in the $J$ band and $\sim$ 56~light-days in the $K$ band. We extend our fits of $\tau \propto \lambda^{4/3}$ to the optical data to estimate the expected near-IR disk lags for these bands. The uncertainty on $\tau_0$ and $y_0$ are used to create the upper and lower bounds for the lags. We extrapolate lags from each of the lag methods. For PyCCF, $J = 17\pm7$~days and $K = 48\pm16$~days. For JAVELIN, $J = 14\pm7$~days and $K = 36\pm14$~days. For PyROA, $J = 17\pm4$~days and $K = 47\pm8$~days. Our extrapolation of the disk lags is in general agreement with what is measured by \cite{Landt2022}.

 We also fit the time lags to the analytical prescription described in \cite{2021ApJ...907...20K} in Fig.~\ref{fig:lags}. Black hole mass and X-ray luminosity are found in literature, leaving the mass accretion fraction ($\dot{m}_E$) and X-ray corona height ($H$) to be fit. We calculated the Eddington fraction using a bolometric luminosity $\dot{m}_E$ = 0.416, so we fit for both when $\dot{m}$ is fixed to this value and when it is allowed to be a free parameter. However, when leaving $\dot{m}_E$ as a free parameter the fit is poorly constrained. We therefore only consider fits with $\dot{m}_E$ fixed at 0.416. Fig.~\ref{fig:lags} only shows the fit for when $\dot{m}_E$ is a fixed parameter, with the red dashed line representing $a^*$ = 0.988 regime. All of the values determined from fitting can be found in Table~\ref{table:alphas}. The values of $H$ are found to be around 30-50 $R_G$, but agree with each other to within 1$\sigma$.

 Spectra of Mrk~876 (Fig.~\ref{fig:spectra_allfilters}) show that there is a significant contribution from the H$\alpha$ broad line that is present in $i$-band measurements. This emission remains significant throughout the duration of the campaign. While determining the exact impact is beyond the scope of this paper, emission from the broad-line region has been suspected to influence lags in prior continuum reverberation mapping campaigns. Contributions from broad emission lines can also influence the lag in a photometric band \citep{2013ApJ...772....9C}. While this is not prominent in all objects, the redshift of Mrk~876 puts the strong H$\alpha$ line in the middle of the $i$ band, indicating that it is a strong possibility in this case.
 
We perform a flux-flux analysis on the dereddened and redshift-corrected flux in Fig.~\ref{fig:fluxflux_0.027}. The values determined for the SED are recorded in Table~\ref{table:fluxflux}.
 The flux-flux analysis allows a determination of the variable and constant components of the SED. The variable (rms) component agrees with a $f_{\nu} \propto \lambda^{-1/3}$ spectrum expected for an accretion disk, though it shows an excess in the $i$ band. This indicates the presence of additional variability beyond what is expected from the accretion disk. The constant spectrum also shows an excess in the $i$ band. Both the variable and constant component excesses can be attributed to a prominent H$\alpha$ line. 
 
 This excess variability lends credence to the broad H$\alpha$ line being the source of the longer than expected $i$-band lag. The H$\alpha$ line has variations smaller than what our detrending removes but longer than what the accretion disk is expected to create. Our detrending process removes slow variations on timescales around 100~days, while the H$\alpha$ in Mrk 876 has been measured to vary on timescales of roughly 43$^{+40}_{-22}$ days \citep{2000ApJ...533..631K}. This allows its variations to influence the $i$-band lags still, despite the detrending process. As explained in Sec.~\ref{sec:spectralanalysis}, we do not expect the lags of H$\alpha$ and the accretion disk to add together simply, but the effect on the lags is clear to see.
 
 We estimate the Eddington fraction using the bolometric luminosity to be $\dot{m}_{\rm E} = 0.416$ during the campaign. This makes it one of the highest Eddington rate AGN studied via continuum reverberation mapping to date. Based on its black hole mass and mass accretion rate, like many other studies have found \citep{2016ApJ...821...56F, 2018ApJ...857...53C, 2019ApJ...870..123E, 2020ApJ...896....1C, 2021ApJ...922..151K}, we find that the normalization ($\tau_0$) recovered from fitting the measured time lags (9.22 $\pm$ 2.64~days, 6.50 $\pm$ 2.00~days, and 9.16 $\pm$~1.38~days for PyCCF, JAVELIN, and PyROA respectively) is several times larger than $\tau_0 = 2.57$~days calculated from theory when $X = 2.54$. These values range from 2.6-3.6 times greater than theory, depending on the method. This implies that a flux-weighted mean radius alone cannot adequately describe the measured accretion disk sizes with the other assumptions about accretion disks we applied. In order for a flux-weighted mean radius model of the accretion disk to recover the size of the disk measured, we would need to assume a much higher accretion rate than expected, higher Eddington ratio, or lower accretion efficiency. This problem exists beyond just reverberation lag measurements, as similar results are found through gravitational microlensing campaigns \citep[e.g.,][]{2008ApJ...673...34P, 2013ApJ...769...53M}. 
 
The additional phenomena required are broadly divided into two categories. First are theories that the source of the lags is still X-ray reprocessing, but that some aspect of AGN geometry is either incorrectly assumed or different than expected. One example of this would be that the irradiating source is higher above the accretion disk than usually assumed \citep{2019ApJ...879L..24K}. The other category of explanations involve different sources for the lags, such as lags instead being due to disk turbulence \citep{2020ApJ...892...63C} or due to far-UV illumination from the inner disk shining onto and providing the reprocessing radiation for the rest of the disk \citep{2017MNRAS.470.3591G}. Alternatively, \cite{Gaskell2018} suggest these longer than expected lags can be attributed to an underestimate of the intrinsic flux of the AGN, and hence an underestimate of the Eddington ratio, due to the presence of large amounts of intrinsic reddening. However, that the variable spectrum closely follows $f_{\nu} \propto \lambda^{-1/3}$ would seem to suggest that there is not a large amount of intrinsic reddening in this particular object.

For our analysis specifically, the choice of detrending via a moving boxcar average could influence the lags and therefore the measured size of the accretion disk. While the majority of detrending lengths we test agree with each other within uncertainty (Fig. \ref{fig:DetrendTimeLagComp}), the outlier cases show that the different lengths produce smaller time lags. We argue that the best detrend length is that of 100 days due to it having the smallest uncertainties, but again this does not guarantee that it is the correct length. The equation we use to calculate the lags (Eq. \ref{eq:alpha}) is also simplistic in its assumption about the geometry of the AGN and accretion disk system.

Another possible explanation for the accretion disk size problem is an underestimation of $X$. Many studies use the value calculated by \cite{2016ApJ...821...56F} of 2.54. However, other studies have calculated it considering other factors of the AGN. When including variation of the disk emission \citep[]{2018MNRAS.473...80T}, the value of $X$ then becomes $5.04$. Using this value, we calculate $\tau_0$~=~6.58~days. This is closer to what we observe for all lag measurement methods, falling within 1$\sigma$ for all methods except PyROA.

Given its mass and Eddington fraction, the continuum lags in Mrk~876 are some of the longest yet observed. The outer edge of the accretion disk is expected to become self-gravitating at 12 light~days regardless of the mass of the system \citep{2022MNRAS.511.1992L}. Our $u$ to $z$ lag is around 13~days, and the measured $\tau_0$ of around 9~days suggests that the $z$ band corresponds to a disk size of 17~-~20~days (depending on lag method), significantly larger than the 12~light~days self-gravitating radius. Our estimates are consistent with what \cite{Landt2022} estimate through spectral fitting.



\section{Conclusions}
\label{sec:Conclusions}
In summary, Mrk~876 displays large amplitude variability over 3~years and shows significant time lags across the optical band. The lags in Mrk~876 are some of the longest continuum lags yet measured, and longer than expected for the self-gravitating radius. Our conclusions on Mrk 876 are as follows:
\begin{itemize}
    \item We measure broad CCFs with the base lightcurves, indicating long term variations are present. The data are detrended by subtracting a moving boxcar average of 100 days, recovering the short term lags. We measure the lags with PyCCF, JAVELIN, and PyROA. The results can be found in Table \ref{table:lags} and plotted in Fig. \ref{fig:lags}.
    \item The $i$-band lag is longer than expected from an extrapolation of the other bands. We analyze spectra taken from before and after the campaign, finding that the H$\alpha$ broad line emission has a strong contribution in the $i$ band -- up about 1/3 of the total $i$-band flux. We find that due to its intermediately long lags ($\sim$40~days) this signal would not be removed by detrending and would still exist in our lag measurements of the detrended data. 
    \item We perform a flux-flux analysis on both the base and detrended data. The base data contains an $i$-band excess in both the constant and variable (rms) emissions, while the rest of the bands agree with the expected profile for an accreting thin disk $f_\nu \propto \lambda^{-1/3}$. The detrended flux-flux analysis reveals that the excess variable emission is removed via detrending, implying that what remains as an excess in the detrended light curves varies on scales longer than reverberation mapping but shorter than 100 days. This adds further support that H$\alpha$ is responsible for this lag excess.
    \item The normalization parameter $\tau_0$ is found for all lag measurement methods. We calculate this value following the parameterization described by \cite{2016ApJ...821...56F}. Two different values of the factor $X$, 2.49 \citep{2016ApJ...821...56F} and 5.04 \citep{2018MNRAS.473...80T}, are used to calculate $\tau_0$ = 2.57~days and $\tau_0$ = 6.58~days, respectively. Our $\tau_0$ values are closer to the latter, agreeing for most methods within uncertainty.
    \item The lags are fit to the analytical prescription described in \cite{2021ApJ...907...20K}. When $\dot{m}_e$ is fixed at the observed value of 0.416, we find an X-ray source height of 30-50 $R_G$ for a maximally-spinning black hole.
\end{itemize}
Continuum reverberation mapping continues to challenge the standard picture of AGN accretion. More studies with high cadence observations are required to truly understand the AGN system.
\newline
\newline
\noindent We thank the anonymous referee for their comments and suggestions. JAM and EMC gratefully acknowledge support from the National Science Foundation through AST-1909199. We thank David Moutard for feedback that has improved this manuscript. This research made use of Photutils, an Astropy \citep{2013A&A...558A..33A, 2018AJ....156..123A} package for detection and photometry of astronomical sources \citep{2021zndo...5796924B}. This work makes use of observations from the Las Cumbres Observatory global telescope network. Research at UC Irvine is supported by NSF grant AST-1907290. HL acknowledges a Daphne Jackson Fellowship sponsored by the Science and Technology Facilities Council (STFC), UK. ERC acknowledges support by the NRF of South Africa. TT acknowledges support from NSF through grant NSF-AST-1907208. The Pan-STARRS1 Surveys (PS1) and the PS1 public science archive have been made possible through contributions by the Institute for Astronomy, the University of Hawaii, the Pan-STARRS Project Office, the Max-Planck Society and its participating institutes, the Max Planck Institute for Astronomy, Heidelberg and the Max Planck Institute for Extraterrestrial Physics, Garching, The Johns Hopkins University, Durham University, the University of Edinburgh, the Queen's University Belfast, the Harvard-Smithsonian Center for Astrophysics, the Las Cumbres Observatory Global Telescope Network Incorporated, the National Central University of Taiwan, the Space Telescope Science Institute, the National Aeronautics and Space Administration under Grant No. NNX08AR22G issued through the Planetary Science Division of the NASA Science Mission Directorate, the National Science Foundation Grant No. AST-1238877, the University of Maryland, Eotvos Lorand University (ELTE), the Los Alamos National Laboratory, and the Gordon and Betty Moore Foundation.

\facilities{LCOGT (elp06, elp08, FTN, optical) \citep{2013PASP..125.1031B}, Liverpool:2m \citep{2004SPIE.5489..679S},  Swift \citep{2004ApJ...611.1005G}, Zowada \citep{2022arXiv220313400C}}
\software{Astropy \citep{2013A&A...558A..33A, 2018AJ....156..123A, 2022ApJ...935..167A}, CALI \citep{2014ApJ...786L...6L}, JAVELIN \citep{2011ApJ...735...80Z, 2016ApJ...819..122Z}, photutils166 \citep{2021zndo...5796924B}, PyCCF \citep{2018ascl.soft05032S}, PyROA \citep{2021MNRAS.508.5449D}}

\bibliography{Mrk876.bib}{}
\bibliographystyle{mnras}



\end{document}